\RequirePackage{fixltx2e}
\documentclass[notitlepage,twocolumn,prl,floatfix,showpacs,aps]{revtex4-1}
\usepackage[nodayofweek]{datetime}
\usepackage{paralist}
\usepackage{amsmath}
\usepackage{mathtools}
\usepackage{natbib}
\usepackage{color}
\usepackage{nicefrac}
\usepackage{hyperref}
\usepackage{graphicx}
\usepackage[caption=false]{subfig}
\newcommand{\kms}{\,\textrm{km s}^{-1}}

\usepackage{multirow}
\usepackage{tabulary}
\usepackage{array}
\usepackage{breakurl}

\begin{document}

\title{Model independent determination of the dark matter mass from direct detection experiments}

\author{Bradley J. Kavanagh}
\email{Electronic address: ppxbk2@nottingham.ac.uk}
\author{Anne M. Green}
\email{Electronic address: anne.green@nottingham.ac.uk}
\affiliation{School of Physics \& Astronomy,University of Nottingham, University Park, Nottingham, NG7 2RD, UK}

\date{\today}

\begin{abstract}
Determining the dark matter (DM) mass is of paramount importance for understanding dark matter. We present a novel parametrization of the DM speed distribution which will allow the DM mass to be accurately measured using data from Weakly Interacting Massive Particle (WIMP) direct detection experiments. Specifically, we parametrize the natural logarithm of the speed distribution as a polynomial in the speed \(v\). We demonstrate, using mock data from upcoming experiments, that by fitting the WIMP mass and interaction cross-section, along with the polynomial coefficients, we can accurately reconstruct both the WIMP mass and speed distribution. This new method is the first demonstration that an accurate, unbiased reconstruction of the WIMP mass is possible without prior assumptions about the distribution function. We anticipate that this technique will be invaluable in the analysis of future experimental data.
\end{abstract}


\maketitle

Dark matter (DM) has thus far only been detected through its gravitational interaction with Standard Model particles and its particle nature is not yet known. Weakly Interacting Massive Particles (WIMPs) are a good DM candidate as they are generically produced in the early Universe with the required abundance and Supersymmetry (SUSY) provides a concrete well-motivated WIMP candidate in the form of the lightest neutralino (e.g. Ref.~\cite{DMO4}). WIMPs can be detected directly in the lab or indirectly via their annihilation products. They can also be produced at particle colliders such as the LHC. Direct detection experiments \cite{DMO11,DMO9} aim to observe nuclear recoils produced by WIMPs passing through terrestrial detectors (such as XENON100 \cite{DMDD112} or CDMS \cite{DMDD63}). While  WIMPs are within the reach of near future direct detection experiments their convincing detection is likely to require consistent signals (i.e. with the same WIMP parameters) from direct, indirect and collider searches (e.g. Ref.~\cite{Bertone:2010at}).
An accurate determination of the WIMP parameters would also be instrumental in constraining the parameter space of SUSY models (e.g. Ref.~\cite{Strege:2012bt}).

Measuring the rate of nuclear recoils as a function of recoil energy should in principle allow the WIMP mass and interaction cross-section with nucleons to be extracted. However, the analysis of direct detection data requires  assumptions to be made about the astrophysical distribution of the DM within the Milky Way halo. The local velocity distribution, \(f(\textbf{v})\), encodes the speeds of DM particles and determines the recoil energies observed in experiments. Direct detection experiments usually assume the simplest possible model for the Milky Way halo (referred to as the \textit{Standard Halo Model}). This model assumes that the halo is isotropic, has a density distribution $\rho(r) \propto r^{-2}$ and has reached a steady state, in which case it has a Maxwell-Boltzmann velocity distribution in the Galactic frame. However, high resolution N-body simulations \cite{DMA43,Vogelsberger:2008qb} suggest that the true distribution function is non-Maxwellian. In particular there may be features in the high-speed tail of the speed distribution from particles that are not yet virialised \cite{Kuhlen:2012fz}. Furthermore, the effect of baryons on the DM halo is not yet fully understood. For example, some simulations \cite{Read:2008fh,DMA54} show evidence for a dark disk (DD) which corotates with the baryonic disk and which may significantly affect the direct detection rate \cite{Bruch:2008rx}.

These uncertainties in the velocity distribution can lead to an order of magnitude uncertainty in estimates of the interaction cross-section \cite{DMDD39} and significant bias in the recovery of the WIMP mass from direct detection data \cite{DMDD3}. It is therefore imperative to account for this uncertainty in the current and future analysis of such experiments. Several different approaches have been proposed. One option is to simultaneously fit Galactic parameters, for example the lag speed (the speed of the Solar System with respect to the peak of the WIMP speed distribution), the velocity dispersion and anisotropy, alongside the WIMP mass and cross-section \cite{DMDD39,Peter:2009ak,DMDD121}. However  this method assumes that the Milky Way halo is fully equilibrated and requires that it can be described by one of a relatively small class of models. Peter proposed a more model-independent approach where \(f(v)\) is parameterized as a series of constant bins in velocity space~\cite{DMDD3}. Subsequent studies have explored parameterizing the momentum distribution \cite{BJK1}, while Ref.~\cite{DMDD85} parameterises \(f(v)\) in terms of integrals of motion. However, these approaches still have significant shortcomings and either result in a substantial bias in WIMP parameters \cite{DMDD3,BJK1} or assume that the WIMP mass is already known \cite{DMDD85}.

In this letter we present a new model-independent parametrization method for \(f(v)\) which allows the accurate, unbiased reconstruction of the WIMP mass. In order to demonstrate the robustness of the method, we generate mock data sets for three proposed experiments (modelled on detectors which are in development) using several benchmark speed distributions.

The event rate \(R\) per unit nuclear recoil energy \(E_R\) in a DM detector can be written as \cite{DMO4}
\begin{equation}
\label{eq:Rate}
\frac{\textrm{d}R}{\textrm{d}E_R} = \frac{\rho_0 \sigma_p}{2 m_\chi \mu_{\chi p}^2} A^2 F^2(E_R) \eta(v_\textrm{min})\,,
\end{equation}
where \(\rho_0\) is the local DM mass density, \(m_\chi\) is the WIMP mass, \(\sigma_p\) is the WIMP-proton cross-section (at zero-momentum transfer) and \(\mu_{\chi p}\) is the WIMP-proton reduced mass, \(\mu_{\chi p} = (m_\chi m_p)/(m_\chi + m_p)\). Here, we have considered only the spin-independent contribution to the rate, which dominates for targets with large atomic mass number, \(A\). Assuming that the coupling to protons and neutrons is identical, this leads to an \(A^2\) enhancement in the scattering rate \cite{DMO4}. Finally, the loss of coherence due to the finite size of the nucleus is accounted for by the form factor, \(F^2(E_R)\), which we take to have the Helm form \cite{Helm}.

The local velocity distribution of Galactic WIMPs enters into the event rate through \(\eta(v_{\textrm{min}})\), the mean inverse speed:

\begin{equation}
\eta(v_{\textrm{min}}) = \int_{v_{\textrm{min}}}^\infty \frac{f(\textbf{v})}{v}\, \textrm{d}^3\textbf{v} \,,
\end{equation}
where \(f(\textbf{v})\) is the normalised velocity distribution in the Earth frame and \(v_{\textrm{min}}\) is the minimum WIMP speed required to produce a recoil of energy \(E_R\). For a detector of nuclear mass, \(m_N\), and reduced WIMP-nucleon mass, \(\mu_{\chi N} = (m_\chi m_N)/(m_\chi + m_N)\), this minimum required speed is

\begin{equation}
v_{\textrm{min}}(E_R, m_\chi, m_N) = \sqrt{\frac{m_N E_R}{2\mu_{\chi N}^2}} \,.
\end{equation}

In order to generate mock data sets, we use three hypothetical experiments which are representative of those currently in development (namely XENON1T \cite{DMDD95}, SuperCDMS \cite{DMDD74} and WArP \cite{DMDD94}) and we assume for concreteness that our proposed detectors have negligible backgrounds and perfect energy resolution. A previous study \cite{DMDD51} found that including an uncertainty of order 1 keV in the measurement of the recoil energy does not affect parameter reconstruction and therefore we expect our results to generalize to finite resolution. Each experiment is then characterised by the mass of the target nucleus \(m_N\), the total detector mass \(m_\textrm{det}\), the effective exposure time \(t_\textrm{exp}\) and energy window accessible to the experiment \(\left[E_\textrm{min}, E_\textrm{max}\right]\). Information about detector efficiency is accounted for in the value of \(t_\textrm{exp}\). The values used for these parameters in the current work are shown in Table \ref{tab:Expts}. These have been chosen to closely match those used in Ref.\ \cite{DMDD3} and \cite{BJK1}, though with increased exposure times.

\begin{table}
  \begin{center}
    \begin{ruledtabular}
	\begin{tabular}{m{3.7 cm} m{1.6 cm} m{1.7 cm} m{1.2 cm}}
	  & XENON1T \cite{DMDD95} & SuperCDMS \cite{DMDD74} & WArP \cite{DMDD94} \\
	  \hline
	  Detector Target & Xe & Ge & Ar \\
	  Nuclear Mass, \(m_\textrm{N} / \textrm{amu}\) & 131 & 73 & 40 \\
	  Detector Mass, \(m_\textrm{det} / \textrm{kg} \) & 1000  & 100 & 1000 \\
	  Exposure Time, \(t_\textrm{exp} / \textrm{days} \) & 152  & 274 & 912\\
	  Energy Range,  & [2,30] & [10,100] & [30,130] \\
	  \(\left[E_\textrm{min},E_\textrm{max} \right] / \textrm{keV}\) & & &  \\
	\end{tabular}
    \end{ruledtabular}
  \end{center}
  \caption{Parameter values for the three mock experiments used in this work. The meanings of the experimental parameters are described in the text.}
  \label{tab:Expts}
\end{table}

\begin{figure}[t]
    \includegraphics[width=0.5\textwidth]{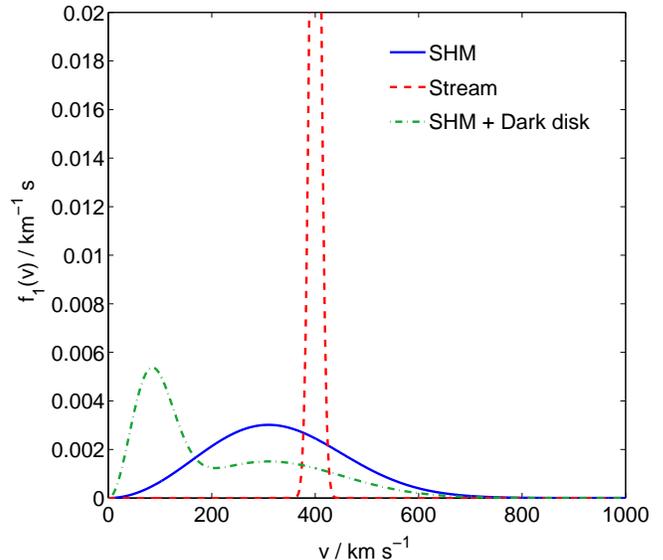}
    \caption{The speed distribution for the three benchmark models defined in the text: SHM (solid blue), SHM+DD (dot-dashed green) and stream (dashed red).}
\label{fig:Distributions}
\end{figure}

We consider three possible underlying distribution functions. The first is the Standard Halo Model (SHM), which we model as a Gaussian in the Galactic frame with width \(\sigma_v = 156 \kms\) and lag speed \(v_{\textrm{lag}} = 230 \kms\). The remaining two extreme distributions lead to significantly different results to the SHM and one would expect them to be difficult to reconstruct. These are: a DM stream (\(v_{\textrm{lag}} = 400 \kms\), \(\sigma_v = 10 \kms\)) and an SHM with a 50\% admixture of a dark disk (\(v_{\textrm{lag}} = \sigma_v = 50 \kms\)). These are referred to as `stream' and `SHM+DD' respectively. All three distributions are shown in Fig.\ \ref{fig:Distributions}.

The experiments used here are typically insensitive to low mass WIMPs (\(m_\chi \lesssim 10 \textrm{ GeV}\)), while higher mass WIMPs (\(m_\chi \gtrsim 100 \textrm{ GeV}\)) lead to a degeneracy in the \(m_\chi-\sigma_p\) plane which has been explored thoroughly in the literature (see Ref.~\cite{DMDD50,DMDD3}, for example). We therefore consider a single benchmark WIMP mass of \(50 \textrm{ GeV}\) in the current work, though we have checked that the conclusions presented here are robust to changes in the benchmark mass. \footnote{See Supplemental Material at \url{http://link.aps.org/supplemental/10.1103/PhysRevLett.111.031302} for an example of a reconstruction using a higher mass benchmark.}

 We use a fiducial cross-section of \(\sigma_p = 2 \times 10^{-45} \textrm{ cm}^2\), which is at the current bound from XENON100 \cite{DMDD112}. However, this bound was calculated assuming a Standard Halo Model and, in any case, the value of \(\sigma_p\) is degenerate with the local DM density \(\rho_0\), which is typically presented with substantial uncertainties (e.g.\ Ref. \cite{DMA13,DMA36,DMA34,DMA48}). We therefore choose to fix the cross-section at the above fiducial value and fix the local DM density at \(\rho_0 = 0.3 \textrm{ GeV cm}^{-3}\). We consider only a single benchmark cross-section, as varying this parameter only affects the total number of events observed and should therefore only affect the precision of reconstructions rather than their accuracy.

 For each set of benchmark parameters we generate a single realisation of data. The impact of Poissonian statistics on reconstructions has previously been studied in detail in Ref.\ \cite{DMDD40}. Using these benchmark values, the total number of events observed across all three experiments ranges from 176 in the SHM+DD case to 386 for the stream.

We assume no prior knowledge of the speed distribution and attempt to construct a parametrization which will allow us to fit the mock data sets in a model independent way. The one (rather weak) assumption we make about the speed distribution is that \(f(v > v_\textrm{max}) = 0\) for \(v_\textrm{max} = 1000 \kms\). The choice of the value \(v_\textrm{max}\) is somewhat arbitrary, but this particular value is conservative as the Galactic escape speed is significantly smaller than this \cite{DMA43}.

We note that, by definition, the speed distribution is everywhere positive \(f(v) \geq 0\), motivating us to parametrize the logarithm of \(f(v)\) as a polynomial of degree \(N\) in \(v\). We therefore write
\begin{equation}
f(v) \propto \exp\left\{ -\sum_{k=0}^N a_k \tilde{P_k}\left(\frac{v}{v_\textrm{max}}\right)  \right\}\,,
\end{equation}
meaning that the full directionally-averaged speed distribution is
\begin{equation}
f_1(v) = v^2 \exp\left\{ -\sum_{k=0}^N a_k \tilde{P_k}\left(\frac{v}{v_\textrm{max}}\right)  \right\}\,,
\end{equation}
subject to the normalisation condition
\begin{equation}
\int_{0}^{v_\textrm{max}} f_1(v) \, \textrm{d}v = 1\,.
\end{equation}

We have chosen a basis of shifted Legendre polynomials \(\tilde{P_k}\) of degree \(k = \{0,1,...,N\}\) for the parametrization. In theory, any basis may be used, but in practice choosing polynomials which are orthogonal over some finite range of \(v\) improves the behaviour of the coefficients \(a_k\). By varying \(N\), the number of terms in the polynomial, we can accommodate features in the distribution function, such as multiple components. By widening the priors on the values \(a_k\), we can accommodate sharper individual structures.

We explore the posterior distribution for the parameter space \(\left\{m_\chi, \sigma_p, \left\{ a_k\right\}\right\}\) using the publicly available \textsc{MultiNest} package \cite{MCMC11,MCMC12} as a generic nested sampler. We use an unbinned likelihood function (as introduced in Ref.~\cite{DMDD90}) and use the following sampling parameters: \texttt{N\(_{\texttt{live}}\) = 10,000}, \texttt{eff = 0.3}, \texttt{tol = 1.0e-4}. We use log-flat priors on \(m_\chi \in [1, 1000] \textrm{ GeV}\) and \(\sigma_p \in [1, 1000] \times 10^{-46} \textrm{ cm}^2\), linearly-flat priors on \(|a_i| < 50\) and go up to degree N = 4 in the basis polynomials. For the stream benchmark, we allow \(|a_i| < 500\) and for the SHM+DD distribution, we extend the basis up to N = 9. We can straightforwardly check the robustness of a given reconstruction by increasing \(N\) and widening the priors on \(a_k\).

\begin{figure}[t]
    \includegraphics[width=0.5\textwidth]{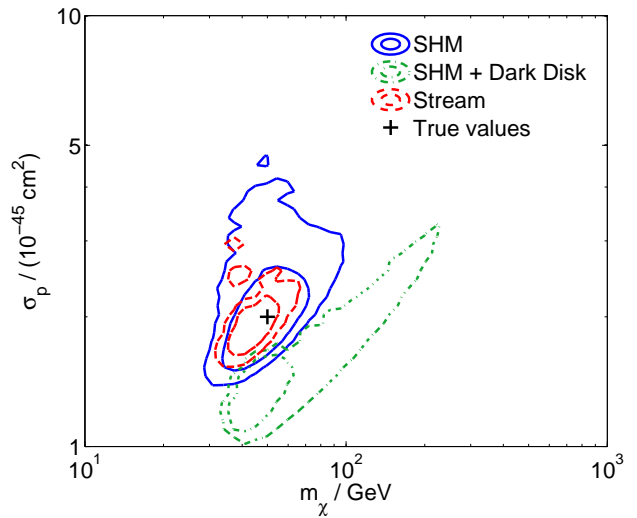}
    \caption{ 68\% and 95\% confidence contours for the WIMP mass \(m_\chi\) and cross-section \(\sigma_p\) obtained using the speed parametrization described in the text. Results are shown for three different underlying speed distributions: SHM (solid blue), SHM+DD (dot-dashed green) and stream (dashed red). The true parameter values are shown as a black cross.}
\label{fig:contours}
\end{figure}

In Fig.\ \ref{fig:contours}, we show the 68\% and 95\% confidence contours for the reconstruction of the mass and cross-section, obtained from the 2-dimensional profile likelihood. In the case of the SHM and stream, the WIMP mass and cross-section are both well recovered, lying within the 68\% contour. In the case of the SHM+DD distribution function, however, there is a significant bias towards a smaller cross-section, with a discrepancy of around 30\% between the underlying and reconstructed cross-section.

 Table~\ref{tab:ReconstructedMass} shows the best fit WIMP masses and error estimates derived from these profile likelihoods. In contrast to the bias observed in the reconstruction of the cross-section, the WIMP mass is accurately reconstructed in all three cases, with the true mass lying within the 1-\(\sigma\) errors. For comparison, Table~\ref{tab:ReconstructedMass} also shows the value of the mass reconstructed using the speed binning method of Peter \cite{DMDD3} and the momentum binning method of Ref.~\cite{BJK1} for the stream benchmark. Both of these methods show a bias towards lower masses and significantly underestimated errors. The new parametrization method presented here is seen to perform significantly better in recovering the WIMP mass.

\begin{table}[t]
  \setlength{\extrarowheight}{3pt}
  \begin{center}
    \begin{ruledtabular}
	\begin{tabular}{lc}
	 Speed distribution & Reconstructed mass (GeV) \\
	 \hline
	 Standard Halo Model (SHM) & \(50.0^{+12.6}_{-9.2}\) \\
	 SHM + Dark disk & \(44.2^{+8.1}_{-3.7}\) \\
	 Stream & \(44.7^{+6.9}_{-3.6}\) \\
	 Stream (using 5 speed bins) & \(29.3^{+0.4}_{-1.0}\) \\
	 Stream (using 5 momentum bins) & \(38.2^{+1.6}_{-2.3}\)
	\end{tabular}
    \end{ruledtabular}
  \end{center}
  \caption{Reconstructed mass calculated from the profile likelihood, using the parametrization presented in this letter. Also shown is the mass reconstructed from the stream distribution using the speed binning \cite{DMDD3} and momentum binning \cite{BJK1} methods for comparison.}
\label{tab:ReconstructedMass}
\end{table}

\begin{figure}[t]
    \includegraphics[width=0.5\textwidth]{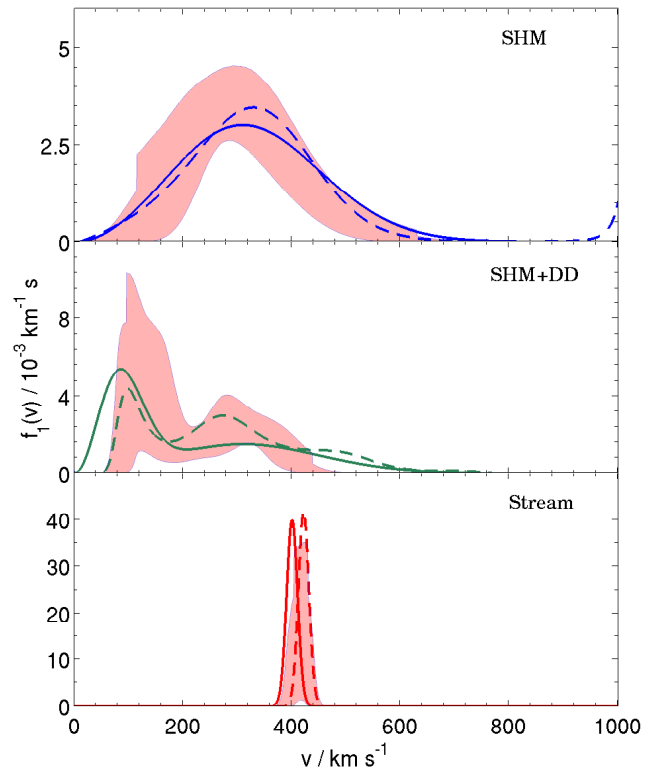}
    \caption{Reconstructed speed distributions for the SHM (top), SHM+DD (middle) and stream (bottom) benchmarks. The true distribution is shown as a solid curve, while the best fit reconstructed value is shown as a dashed curve. Also shown are the marginalised 68\% credible intervals (shaded pink).}
\label{fig:ReconstructedSpeed}
\end{figure}

In Fig.~\ref{fig:ReconstructedSpeed}, we show the reconstructed speed distributions for the three benchmarks. The 68\% credible regions (shaded pink) are obtained by marginalising over the values of \(f(v)\) - derived from the posterior distribution of the \(\left\{a_k\right\}\) - at each value of \(v\). The errors in \(f(v)\) across different speeds are therefore strongly correlated. However, such reconstructions should be representative of the underlying form of the distribution (solid line). Also shown is the best fit form of \(f(v)\) in each case (dashed line).

For all three benchmarks, the speed distribution is largely well recovered. The SHM benchmark recovers a broad distribution, while the 50\% dark disk produces a distinctive double peak structure. The stream benchmark results in a sharp localised peak within \(\sim 20 \kms\) of the true peak. However, the reconstruction is typically poorer for low and high speeds. This is particularly noticeable in the SHM+DD benchmark for \(v < 100 \kms\). This is due to the fact that for a WIMP of mass \(50 \textrm{ GeV}\) and the energy thresholds given in Table~\ref{tab:Expts}, the smallest speed to which the detectors are sensitive is \(v \sim 92 \kms\). Below this speed, the distribution function is poorly constrained. Similarly, at high \(v\), the upper energy limit \(E_\textrm{max}\), combined with form factor suppression of the event rate, leads to low sensitivity.

The three hypothetical experiments probe distinct but overlapping regions of the speed distribution. The high nuclear mass and low energy threshold of XENON1T mean that it probes lower WIMP speeds than the other experiments. Similarly, SuperCDMS probes intermediate speeds and WArP probes high speeds. \footnote{See Supplemental Material at \url{http://link.aps.org/supplemental/10.1103/PhysRevLett.111.031302} for reconstruction results using only pairs of detectors.}.  However, because of the finite energy window of the experiments, we cannot constrain \(f(v)\) across its entire range and therefore we do not know what fraction of DM particles lie within reach of the experiments. Generically then, it is not possible to reconstruct \(\sigma_p\), and the bias in \(\sigma_p\) shown in Fig.\ \ref{fig:contours} is a problem inherent to all model independent approaches. In the case of the SHM+DD benchmark, only 72\% of WIMPs lie within the sensitivity range of the experiments, leading to a reduction of around 30\% in the effective cross-section. In the case of the SHM and stream benchmarks, almost all WIMPs are accessible to the experiments, leading to an accurate reconstruction. However, we do not know \textit{a priori} if this is the case and therefore it is only possible to obtain a lower bound on the cross-section. Nonetheless, this ignorance of the cross-section does not preclude us from accurately reconstructing the DM mass, as we have seen.

In summary, accurately measuring the WIMP mass from direct detection data is crucial in order to convincingly detect, and characterise, DM. We have presented a new parametrization of the DM speed distribution which does so in a model independent way. While this leads to an unavoidable uncertainty in the cross-section, it does allow the speed distributions to be well constrained within the range of sensitivity of the experiments. Most importantly, however, the new parametrization is able to robustly and accurately recover the WIMP mass, even for non-trivial underlying speed distributions. This method demonstrates for the first time that both the WIMP mass and speed distribution can be probed accurately and will therefore be invaluable for the analysis of future experimental data.

\begin{acknowledgments}
The authors thank M. Fornasa and A. J. Christopherson for useful comments. BJK and AMG are both supported by STFC. AMG also acknowledges support from the Leverhulme Trust.
\end{acknowledgments}

\sloppy

\bibliographystyle{apsrev4-1}
\bibliography{bradkav}

\end{document}